\documentclass[%
reprint,
superscriptaddress,
 amsmath,amssymb,
 aps,
 prl,
]{revtex4-2}

\usepackage{xcolor}
\usepackage{graphicx}
\usepackage{bm}
\usepackage{hyperref}
\hypersetup{
colorlinks = true,
linkcolor = blue,
citecolor = blue,
urlcolor = blue}
\begin{document}

\title{Theory of Spin-Dependent Electron Transfer Dynamics \\ at Ar/Co(0001) and Ar/Fe(110) Interfaces}

\author{Moritz M\"uller}
\email{mllr.mrtz@gmail.com}
\affiliation{Donostia International Physics Center (DIPC), Paseo de Manuel Lardizabal 4, San Sebasti\'an-Donostia 20018, Spain}
\affiliation{Centro de F\'isica de Materiales CFM-MPC (CSIC-UPV/EHU), Paseo de Manuel Lardizabal 5, San Sebasti\'an-Donostia 20018, Spain}
\affiliation{CIC nanoGUNE, Tolosa Hiribidea, 76, San Sebasti\'an-Donostia, 20018, Spain}

\author{Pedro Miguel Echenique}
\affiliation{Donostia International Physics Center (DIPC), Paseo de Manuel Lardizabal 4, San Sebasti\'an-Donostia 20018, Spain}
\affiliation{Centro de F\'isica de Materiales CFM-MPC (CSIC-UPV/EHU), Paseo de Manuel Lardizabal 5, San Sebasti\'an-Donostia 20018, Spain}
\affiliation{Departamento de F\'isica de Materiales, Facultad de Ciencias Qu\'imicas, Universidad del Pa\'is Vasco (UPV-EHU), Apdo. 1072, San Sebasti\'an-Donostia 20080, Spain}

\author{Daniel S\'anchez-Portal}
\email{daniel.sanchez@ehu.eus}
\affiliation{Donostia International Physics Center (DIPC), Paseo de Manuel Lardizabal 4, San Sebasti\'an-Donostia 20018, Spain}
\affiliation{Centro de F\'isica de Materiales CFM-MPC (CSIC-UPV/EHU), Paseo de Manuel Lardizabal 5, San Sebasti\'an-Donostia 20018, Spain}

\date{\today}

\begin{abstract}
Recent core-hole-clock experiments [\href{http://link.aps.org/doi/10.1103/PhysRevLett.112.086801}{Phys.\;Rev.\;Lett.\;\textbf{112},\;086801\;(2014)}] showed that the spin dependence of electron injection times at Ar/Co(0001) and Ar/Fe(110) interfaces is at variance with the expectations based on previous calculations for related systems.  Here we reconcile theory and experiment, and demonstrate that the observed dependence is rooted in the details of the spin-split surface band structures. Our {\it ab initio} calculations back that minority electrons are injected significantly faster than majority electrons in line with the experimentally reported ultrashort injection times. The dynamics is particularly sensitive to the size (in reciprocal-space) of the projected band gaps around $\overline{\Gamma}$ for both substrates at the resonance energies. A simple tunneling model incorporating the spin-dependent gap sizes further supports these findings.  
\end{abstract}

\maketitle

Electrons carry charge and spin. This concept  is continuously transforming the field of electronics towards a spin-based discipline known as spintronics. Its applications promise  nonvolatile data storage as well as lossless and ultrafast transmission of information via the spin degree of freedom~\cite{Chappert2007,Hoffmann2015,Hellman2017}. 

Along these lines, the spin-dependence of electron transfer across interfaces must be understood in detail. While advances in free-electron lasing yield new opportunities to directly resolve such ultrafast processes in time~\cite{Hoffmann2015}, current experiments based on the core-hole-clock technique~\cite{Wurth2000,Bruhwiler2002,Bjorneholm1992} are readily able to access spin-resolved charge transfer times down to the sub-femtosecond domain~\citep{Fohlisch2005,Kuhn2019} through spin selective excitation~\cite{Blobner2014} or detection~\cite{Feulner2015,Sundermann2016} in the energy domain.

In particular, recent core-hole-clock experiments by \citet{Blobner2014} probed the spin-dependent dynamics of electron transfer from core-excited Argon atoms towards ferromagnetic Co(0001) and Fe(110) substrates on which they are adsorbed. This study revealed a significant spin dependence of the ultrashort time-scales of the injection process, with the transfer of minority-spin electrons significantly faster than that of majority spin. 

Interestingly, previous theoretical calculations predicted the reverse spin dependence for atomic Cs adsorbates on Fe(110)~\cite{Muino2011}. This behavior was explained in terms of the different character (and, thus, decay into vacuum) of the electronic acceptor states available in the substrate at the relevant energy: dispersive sp-bands for majority spin versus localized d-bands for minority. Although similar correlations were explored in the case of Ar/Co(0001) and Ar/Fe(110)~\cite{Blobner2014}, the ultimate reason regarding the experimentally observed trend remains an open question.

In this letter we explore spin-dependent electron injection from core-excited Argon towards ferromagnetic Co(0001) and Fe(110) surfaces by means of a combination of density functional theory (DFT) calculations and Green's function techniques. 
We find that the first Ar$^*$-resonance above the Fermi level (4s) shows faster charge transfer times for minority than for majority spin paralleling the core-hole-clock experiments by \citet{Blobner2014}.  
The analysis of our data reveals that the size of the electronic gaps around the $\overline{\Gamma}$-point in the surface projected band structures determines this behavior. A simple model relying on minimal ingredients and incorporating the sizes of the band gaps confirms this observation.   

We model core-excited Ar$^*$ on Co(0001) and Fe(110) surfaces with the \textsc{Siesta} code~\cite{Soler2002}, Troullier-Martins-type pseudopotentials~\cite{Troullier1991}, the PBE\;functional~\cite{Perdew1996}, and a double-$\zeta$ polarized basis generated with an energy shift of $0.1$\;eV (the Ar$^*$ basis includes double-$\zeta$ 4s and 3p orbitals and single-$\zeta$ 3d and 4p shells of polarization orbitals). We use slabs containing 11 metal layers separated by $\sim\!40$\;\r{A} of vacuum and relax the outermost layer on each side. Ar atoms are then placed, at fixed distances from the surface, symmetrically on both sides of the slab within a 4$\times$4 lateral supercell. We checked that the results are nearly independent on the adsorption site (see Supplemental Material~\cite{Supplemental}) and here we only present those for top positions. The lattice parameters are $a = 2.88$\;\r{A} for Fe and $a = 2.51$\;\r{A} and $c = 4.09$\;\r{A} for Co. The computational settings include a mesh cutoff of $250$\;Ry, a 5$\times$5$\times$1 and a 6$\times$6$\times$1 Monkhorst-Pack $\bm{k}$-point grid to sample the Fe and Co supercells, respectively.

We aim at computing the resonance spectrum\;$\varrho_\mathrm{R}(E)$ from the projection of the Green's function\;$G(E)$ onto the resonance wave packet\;$\phi_\mathrm{R}$,
\begin{equation}\label{eq:spectrum}
\varrho_\mathrm{R}(E)=-\frac{1}{\pi}\operatorname{Im}\langle\phi_\mathrm{R}|G(E)|\phi_\mathrm{R}\rangle.
\end{equation}
Here, we consider that the resonance wave packet $|\phi_\mathrm{R}\rangle=c_\mathrm{s}|\varphi_\mathrm{4s}\rangle+c_\mathrm{p}|\varphi_\mathrm{4p_\mathrm{z}}\rangle$ is localized on the adsorbate and consists of a linear combination of the 4s- and a polarizing 4p$_\mathrm{z}$-orbital of free Ar$^*$. The amount of polarization was determined employing an optimization scheme aiming at maximizing the area below the resonance peak while keeping it localized in energy (see Supplemental Material~\cite{Supplemental}). Analyzing the resonance spectra we extract the linewidth $\Gamma$ of the 4s-Ar$^*$ resonance and relate it to the mean charge injection time $\tau=\hbar/\Gamma$. 

In order to mimic the excitations produced in the relevant core-hole-clock experiments~\cite{Blobner2014}, we simulate constrained Ar$^*$ atoms including a 2p-hole in the pseudopotential and adding the density corresponding to a spin-polarized electron occupying the 4s-orbital.

Generally, commonly used slab calculations with a finite amount of layers display an energy spacing between discrete sub-bands of  $\sim\!\!\pi^2\hbar^2/(2m^*L^2)$, i.e., $\sim\!90$\;meV for a slab thicknesses of $L\!\approx\!20$\;\r{A} corresponding to $11$~layers of Co(0001) or Fe(110). This energy spacing critically affects the shape of the Ar$^*$ resonances in our simulations, since it is of the order of the linewidths defining the charge transfer times we study. 

To overcome the limitations of such finite slabs, we adopt a recursive Green's function scheme~\cite{SanchezPortal2007,SanchezPortal2007b} to model surfaces consisting of an infinite amount of layers using the \textsc{Transiesta} code~\cite{Brandbyge2002,Papior2017}. In particular, we compute the Green's function\;$G(E)$ of the first 8 Co(0001) layers [6 Fe(110) layers] of which the last 4 (3) layers connect via the self-energy\;$\Sigma(E)$ to an infinite substrate
\begin{equation}
G(E)=[(E+\mathrm{i}\eta)S-H-\Sigma(E)]^{-1},
\end{equation}
where $H$ is the Hamiltonian and $S$ the overlap matrix of the surface layers in our local basis. We use a small shift $\eta=5$\;meV along the complex axis to avoid singularities.

\begin{figure}[t]
\includegraphics[width=0.8\linewidth]{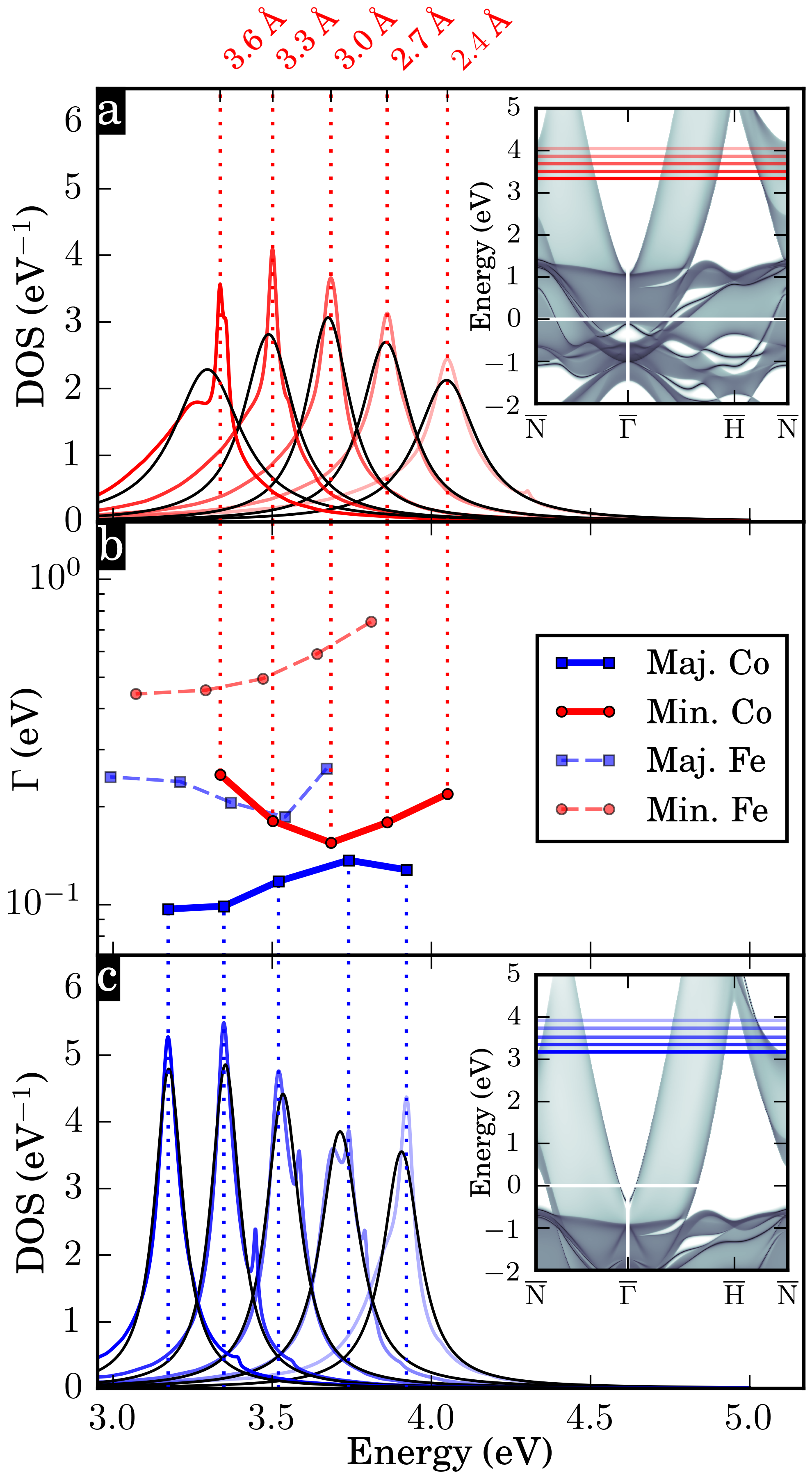}
\caption{\label{fig:1} Computed first resonance above the Fermi level for Ar$^*$(2p$^{-1}$4s) on Co(0001) for adsorption distances ranging from 2.4 to 3.6\;\r{A} in the minority\;(a) and majority spin channel\;(c). Different shadings of the colored lines in the plots encode the adsorption heights. 
The energy positions of the resonance maxima are marked by horizontal lines with respect to the band structure of the Co(0001) surface in the insets, and strongly overlap with the region covered by a band gap around the $\overline{\Gamma}$-point. Panel (b) collects the linewidths $\Gamma$ of the Ar$^*$ resonances according to a  Lorentzian fitting of the peaks (black lines) in (a, c). The spin-dependent linewidths of similar Ar$^*$-resonances on Fe(110) are also shown in (b).}
\end{figure}

Fig.\;\ref{fig:1} shows the calculated resonance spectra (cf. Eq.\;\ref{eq:spectrum}) of the 4s Ar$^*$ resonance for minority (a) and majority spin (c) on Co(0001) as a function of the adsorption height. Here, we consider $k$-averaged spectra since we aim at modeling an isolated adsorbate rather than a periodic array of adsorbates~\cite{Fratesi2014,Muller2018}. 
The peaks shift up in energy as the adsorbate approaches the surface displaying a linear behavior.
We determine the linewidths $\Gamma$ by fitting a Lorentzian $\Omega\pi^{-1}[(E-E_\mathrm{R})^2-\Omega^2]^{-1}$ and compare the results for the two spin-polarizations in Fig.\;\ref{fig:1}b. Note that the small numerical broadening $\eta$ must be subtracted and, thus, $\Gamma=2(\Omega-\eta)$. The plot shows also the case of the Fe(110) substrate. For both substrate materials the extracted linewidths in the minority channel are consistently larger than in the majority channel. 

\begin{figure}[b]
\includegraphics[width=0.8\linewidth]{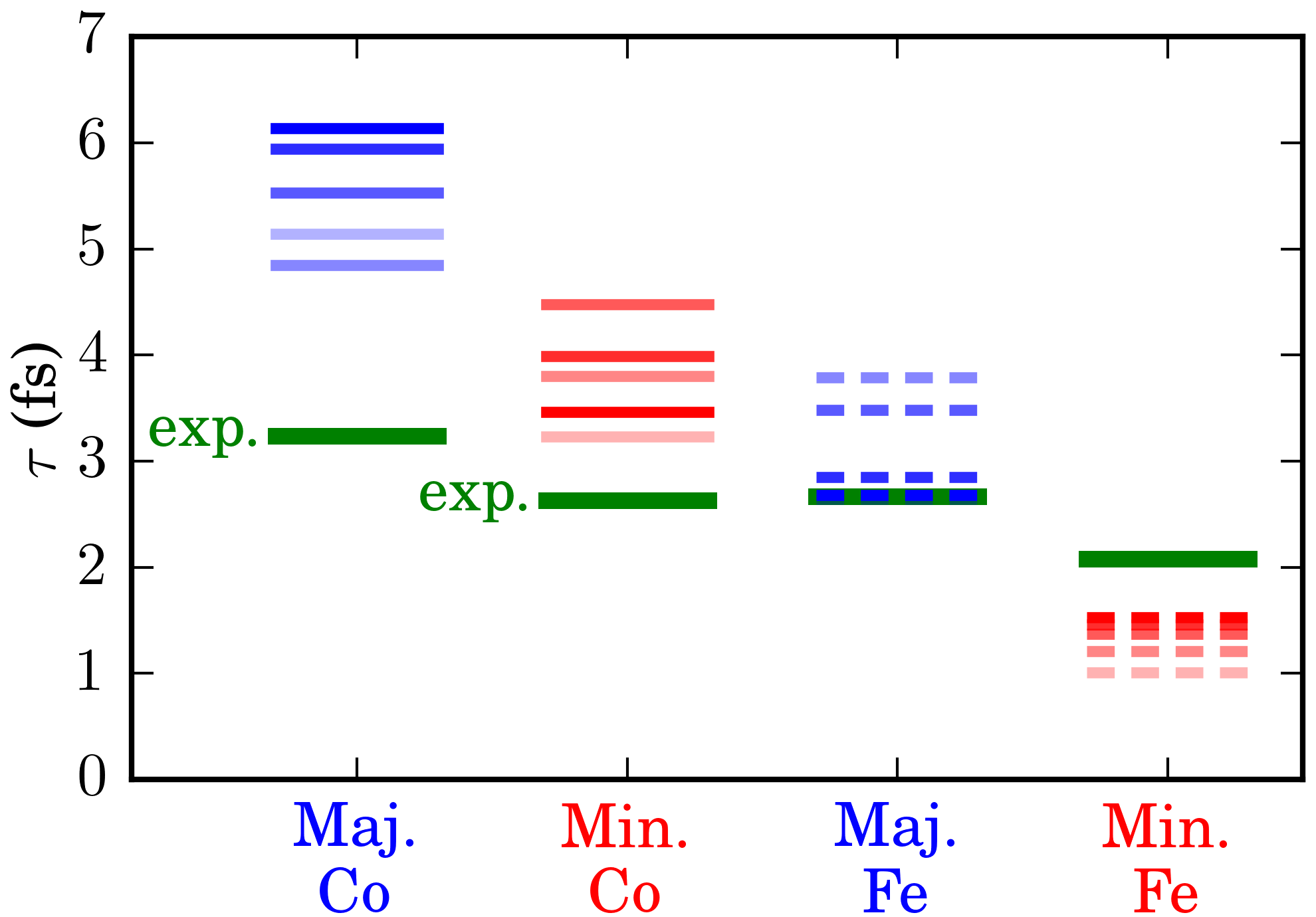}\\
\caption{\label{fig:2} 4s Ar$^*$ lifetimes $\hbar/\Gamma$ derived from the 
linewidths $\Gamma$ in Fig.\;\ref{fig:1}b for minority and majority spin on Co(0001) and Fe(110) (blue and red lines as in Fig.\;\ref{fig:1}) in comparison with experiment~\cite{Blobner2014} (green bars).}
\end{figure}

Translating to electron injection times $\hbar/\Gamma$ we compare our simulations with the experiment~\cite{Blobner2014} in Fig.\;\ref{fig:2}. The time scales are in the right ballpark and show the correct trends: Charge injection from minority channels is faster than from majority channels for each substrate, while the injection time of the minority channel on Co(0001) equals that of the majority channel on Fe(110). Furthermore, we checked (see Supplemental Material~\cite{Supplemental}) that, also in agreement with experiment~\cite{Blobner2014}, in the case of Ni(111) both spin channels differ by less than 15\%. This confirms the validity of the approach and allows for a detailed analysis in order to unveil the origin of the observed effect.

Inspecting the alignment of the resonance positions with the band structure of the surfaces in the insets of Fig.\;\ref{fig:1}a,\;c [only Co(0001) shown, see Supplemental Material for Fe(110)~\cite{Supplemental}] reveals that the resonances are located in regions with a prominent gap in the projected band structure  around $\overline{\Gamma}$. 
Inside the electronic gaps no acceptor states are available so that charge transfer is effectively suppressed. For this reason the size of the projected gaps is a decisive factor controlling the electron dynamics at interfaces~\cite{Gauyacq2004a,Hecht2000,Borisov2001,Borisov1999,Chulkov1997}. Resonances of majority spin, residing deeper inside the electronic gap, bear smaller linewidths than resonance peaks of minority spin, explaining the experimentally observed differences in charge transfer time~\cite{Blobner2014}.

As Ar$^*$ approaches the surface, the presence of gaps in the projected band structure leads to two competing effects: (i) the growing overlap of the Ar$^*$\;states with the wave functions in the substrate increases the linewidth [e.g., minority spin on Fe(110), Fig.\;\ref{fig:1}b];  (ii) the shifting of the resonance peak to higher energies and, thus, deeper into the electronic gaps causes a decrease in the linewidth [e.g., majority spin on Co(0001) for adsorption distances from $3.6$\;\r{A} to $3.0$\;\r{A}, Fig.\;\ref{fig:1}b].

\begin{figure}[b]
\includegraphics[width=0.8\linewidth]{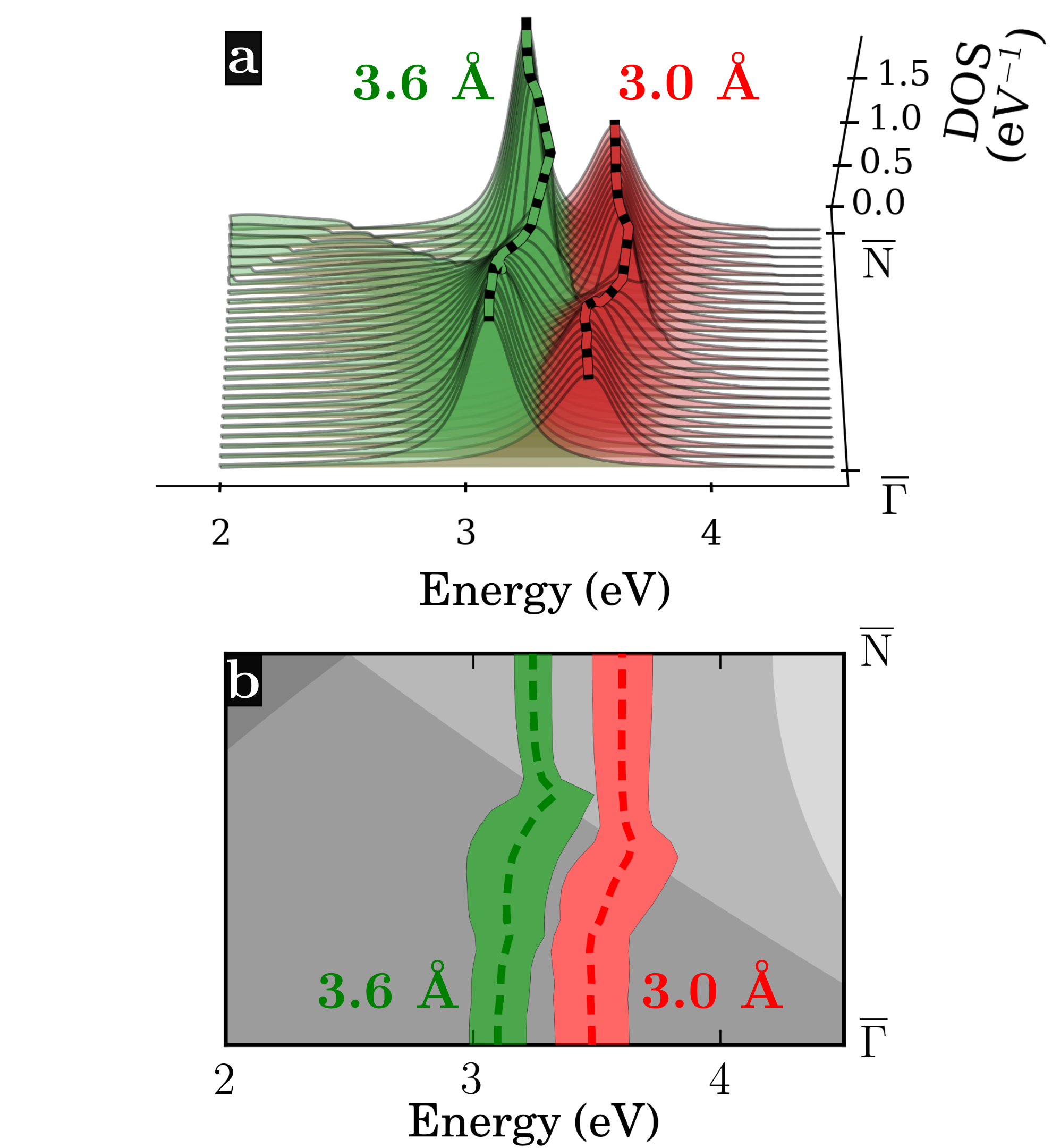}
\caption{\label{fig:3} $k$-dependent minority 4s Ar$^*$-resonance along the $\overline{\Gamma}$---$\overline{\mathrm{N}}$ line on Co(0001) for adsorption distances of 3.6\;\r{A} (green) and 3.0\;\r{A} (red) in a 4$\times$4-supercell (a). 
Shaded areas in (b) show the effect of the folding of the projected band gap around $\overline{\Gamma}$ (using the model band-structure in Fig.\;\ref{fig:4}b), revealing its interplay with the Ar$^*$-resonance. As the adsorbate approaches the surface, the resonance (dashed lines) moves deeper into the electronic gap as reflected by a lower DOS in lighter shades of gray.
In these regions the $k$-dependent widths $\Gamma(k_\parallel)$ --depicted by the red and green areas around the peak positions-- drop in comparison with regions with a larger DOS.}
\end{figure}

Fig.\;\ref{fig:3}a illustrates the interaction of the Ar$^*$4s-resonance with the electronic gap along the $\overline{\Gamma}$---$\overline{\mathrm{N}}$ line  in reciprocal space. Reducing the adsorption distance from $3.6$\;\r{A} to $3.0$\;\r{A} the resonance shifts up in energy, where the size of the electronic gap increases (peak positions marked by dashed lines in Fig.\;\ref{fig:3}a,\;b). 
The folding of the band gap that appears around $\overline{\Gamma}$ of the original surface Brillouin zone gives rise to a region of lower density of states (DOS) around $\overline{\mathrm{N}}$ in the 4$\times$4-supercell (lighter shaded areas represent lower substrate DOS in Fig.\;\ref{fig:3}b). Inside this gap region the extracted $k$-dependent linewidths (from Lorentzian fits) drop as can be seen from the widths of the colored areas in Fig.\;\ref{fig:3}\;b.

\begin{figure}[t]
\includegraphics[width=0.8\linewidth]{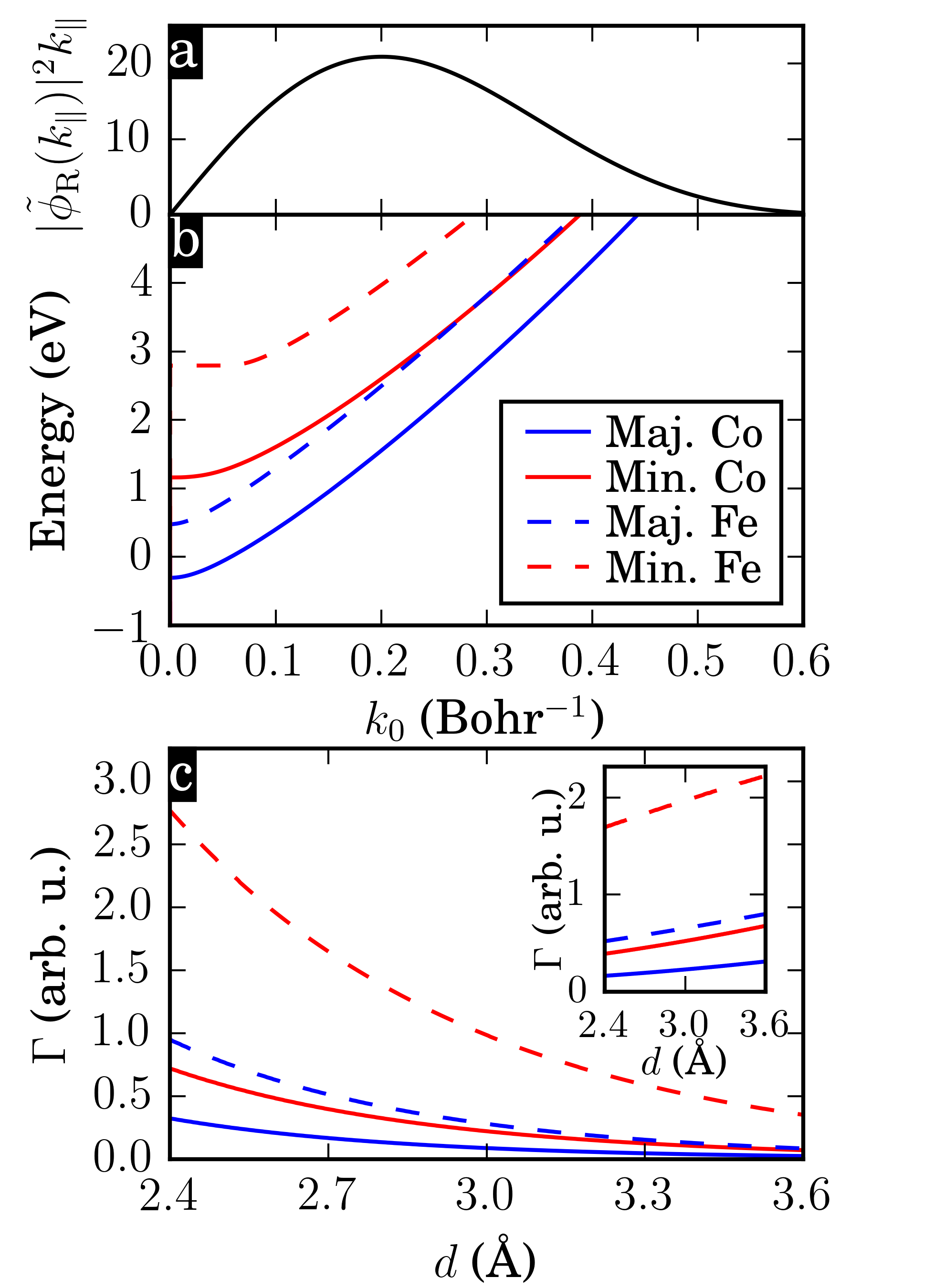}
\caption{\label{fig:4} A simple model incorporating the  spread of the initial wave-packet in reciprocal-space (a), and the energy and spin dependence of the edges of the projected band gap around $\overline{\Gamma}$ (b) yields linewidths in coincidence with the experimental trends~\cite{Blobner2014} regarding spin and material (c). Here, we used $d_0=3$\;Bohr, $\Phi=5$\;eV, and $m^*_\parallel=0.4$. The inset in (c) shows that an unintuitive decrease in tunneling rate is obtained upon lowering the distance $d$ when fixing $(d-d_0)$ to a constant value (here $2$\;Bohr) in Eq.\;\ref{eq:factor} considering $z=d$.} 
\end{figure}

In order to back our observations we employ a simple model that captures the effect of the electronic gap at the relevant energies. In particular, we only consider the dispersive bands in the substrate giving rise to a constant DOS $\rho(E,k_\parallel)\approx C$ outside a disk-like gap region ($k\leq k_0$) with vanishing DOS around the $\overline{\Gamma}$-point. The average tunneling rate is then given by
\begin{equation}\label{eq:coupling}
\Gamma\approx4\pi^2 C\int_{k_0}^{\infty}|V(k_\parallel)|^2 k_\parallel \,\mathrm{d}k_\parallel
\end{equation}
where $V(k_\parallel)$ is the hopping matrix element between the resonance wave packet $\phi_\mathrm{R}(\bm{r})$ and each of the delocalized electronic states $\psi_{\bm{k}_\parallel}(\bm{r})$ with the same energy in the substrate. 
We estimate $V(k)$ to lowest order by their overlap, so that
\begin{alignat}{2}
\label{eq:fourier}
V(\bm{k}_\parallel)
&\approx{\begingroup\textstyle\int\endgroup}\psi_{\bm{k}_\parallel}(\bm{r}) \phi_\mathrm{R}(\bm{r})\,\mathrm{d}\bm{r} \nonumber \\
&\approx{\begingroup\textstyle\int\endgroup}f(z,\bm{k}_\parallel)\mathrm{e}^{-\mathrm{i}\bm{k}_\parallel \bm{r}_\parallel} \phi_\mathrm{R}(\bm{r})\,\mathrm{d}\bm{r}_\parallel\,\mathrm{d}z \\ 
\nonumber
&\approx f(d,\bm{k}_\parallel)\;  {\begingroup\textstyle\int\endgroup}\mathrm{e}^{-\mathrm{i}\bm{k}_\parallel \bm{r}} \phi_\mathrm{R}(\bm{r})\,\mathrm{d}\bm{r} = f(d,\bm{k}_\parallel)\;
\tilde{\phi}_\mathrm{R}(\bm{k}_\parallel). 
\end{alignat}
Here, we factor the wave functions of the substrate $\psi_{\bm{k}_\parallel}(\bm{r})$ into a contribution 
\begin{equation}\label{eq:factor}
f(z,k_\parallel)=\exp\left[-(z\!-\!d_0)\sqrt{2[\Phi-E_\mathrm{R}]+{k_\parallel^2}/{m^*_\parallel}}\;\right],
\end{equation} 
that decays exponentially into vacuum with the distance $z$ to the surface, and a plane wave contribution $\mathrm{e}^{-\mathrm{i}\bm{k}_\parallel \bm{r}_\parallel}$ reflecting the dispersive character of the bands close to the resonance energies $E_\mathrm{R}$. We further simplify the expression by evaluating  $f(z,k_\parallel)$ at the adsorbate's position\;$d$. Finally, the coupling is determined by the product of the Fourier transform of the initial wave packet $\tilde{\phi}_\mathrm{R}(k_\parallel)$ with $f(d,k_\parallel)$. For electrons with large momenta~$k_\parallel$ parallel to the surface the injection probability decreases exponentially (see Eq.\;\ref{eq:factor}): At a given energy $E_R$ the effective injection barrier grows with the kinetic energy\;$k_\parallel^2/{(2m^*_\parallel)}$ of the acceptor state ($m^*_\parallel\approx 0.4$ for both materials). 

The Fourier transform $\tilde{\phi}_\mathrm{R}(k_\parallel)$ also reduces efficiently the coupling to states with large $k_\parallel$. This can be seen in Fig.\;\ref{fig:4}a, where we used the 4s pseudo wave-function~\cite{Troullier1991} of Ar$^*$ as a simple model for the wave packet. The spread in reciprocal space of the resonance wave-packet is comparable to the extension of the projected band gap around $\overline{\Gamma}$ (Fig.\;\ref{fig:4}b), explaining the large impact of this gap on the charge transfer dynamics. Furthermore, at a given energy $E_R$ the size of the projected band gap is considerably larger for majority spin, explaining the larger charge-transfer times for the majority spin channel on both substrates. 

The results of the simple model sketched above are plotted as a function of the distance\;$d$ in Fig.\;\ref{fig:4}c. It is important to notice here that $k_0$ (the lower integration limit in Eq.\;\ref{eq:coupling}) is a function of the resonance energy $E_R$ (Fig.\;\ref{fig:4}b), while $E_R$ depends approximately linearly on~$d$ (Fig.\;\ref{fig:1}). Thus, the linewidths depend on $d$ in several ways. However, for any arbitrary fixed value of $d$ the results of the model reproduce nicely the behavior found in the experiment~\cite{Blobner2014}: majority spin has a larger lifetime than minority spin on both substrates, while majority lifetimes on Fe are similar to those of minority electrons on Co. This confirms the role played by the projected band gap as the determining factor. 

The assumption of an exponential tunneling-like behavior in Eq.\;\ref{eq:factor} leads in all cases to a pronounced decay of the linewidth with distance, which does not fully reproduce the various trends of our {\it ab initio} calculations. In particular, the counter-intuitive decrease of the electron injection rates $\Gamma$ upon approaching the surface (Fig.\;\ref{fig:1}b) is missing. However, this behaviour can be recovered by setting $(d-d_0)$ to a fixed value in Eq.\;\ref{eq:factor} with $z=d$. In such case, the distance dependence is determined by the increase of size of the projected band gap as the resonance shifts up in energy when $d$ is reduced. The result is shown in the inset of Fig.\;\ref{fig:4}c and demonstrates that the projected band gap also plays a key role to determine the distance dependence of resonance linewidths.

In conclusion, we showed that the details of the surface projected band structure and the level alignment are essential to explain the ultra-short time-scales of charge injection and their variations as a function of electron spin and electronic coupling across the interface. In particular, the presence of electronic gaps around  $\overline{\Gamma}$, as found in ferromagnetic Co(0001) and Fe(110) substrates, constitutes an efficient blocking mechanism for the tunneling of electrons and is instrumental to explain the material, spin and energy dependence of electron injection. With this main ingredient we can rationalize and reproduce recent experimental results for electron transfer times from core-excited Ar$^*$ atoms~\cite{Blobner2014}. In general, charge transfer from adsorbates towards surfaces is governed by a combination of the $k$-dependent coupling matrix elements, accounting for the symmetry and spatial overlap of the involved states, and the distribution of the available acceptor states in reciprocal space. These findings emphasize that a detailed understanding of the electronic and atomic structure of the system is a necessary ingredient of any method to simulate ultra-fast dynamics accurately down to the fundamental time-scales of electronic motion. 

\begin{acknowledgments}
We wish to acknowledge illuminating discussions with Prof. U. Heinzmann and Prof. P. Feulner, as well as financial support from the EU FP7 programme under Grant Agreement No. 607232 (THINFACE), the Spanish Ministerio de Ciencia, Innovaci\'on y Universidades (Grant. No. MAT2016-78293-C6-4-R) and the Basque Dep. de Educaci\'on and the UPV/EHU (Grant No. IT1246-19).
\end{acknowledgments}

\bibliography{manuscript}

\end{document}

\title{Supplemental Material: \\ Theory of Spin-Dependent Electron Transfer Dynamics \\ at Ar/Co(0001) and Ar/Fe(110) Interfaces}

\author{Moritz M\"uller}
\email{mllr.mrtz@gmail.com}
\affiliation{Donostia International Physics Center (DIPC), Paseo de Manuel Lardizabal 4, San Sebasti\'an-Donostia 20018, Spain}
\affiliation{Centro de F\'isica de Materiales CFM-MPC (CSIC-UPV/EHU), Paseo de Manuel Lardizabal 5, San Sebasti\'an-Donostia 20018, Spain}
\affiliation{CIC nanoGUNE, Tolosa Hiribidea, 76, San Sebasti\'an-Donostia, 20018, Spain}

\author{Pedro Miguel Echenique}
\affiliation{Donostia International Physics Center (DIPC), Paseo de Manuel Lardizabal 4, San Sebasti\'an-Donostia 20018, Spain}
\affiliation{Centro de F\'isica de Materiales CFM-MPC (CSIC-UPV/EHU), Paseo de Manuel Lardizabal 5, San Sebasti\'an-Donostia 20018, Spain}
\affiliation{Departamento de F\'isica de Materiales, Facultad de Ciencias Qu\'imicas, Universidad del Pa\'is Vasco (UPV-EHU), Apdo. 1072, San Sebasti\'an-Donostia 20080, Spain}

\author{Daniel S\'anchez-Portal}
\email{daniel.sanchez@ehu.eus}
\affiliation{Donostia International Physics Center (DIPC), Paseo de Manuel Lardizabal 4, San Sebasti\'an-Donostia 20018, Spain}
\affiliation{Centro de F\'isica de Materiales CFM-MPC (CSIC-UPV/EHU), Paseo de Manuel Lardizabal 5, San Sebasti\'an-Donostia 20018, Spain}

\date{\today}

\maketitle

\section{Role of the Description of the Resonance Wave Packet: \\ Polarization in the Vicinity of the Surface}
In order to account for a possible polarization of the Ar$^*$-resonance wave packet in the vicinity of the surface we mixed different components of the basis set. In particular we optimized wave packets constructed from Ar$^*$4s- and polarizing Ar$^*$4p$_\textrm{z}$-orbitals of \textsc{siesta}'s numerical atomic orbital basis.

In the following we derive the coefficients $c_\mathrm{0}$~and~$c_\mathrm{1}$ of such wave packets\;$\ket*{\phi_\mathrm{R}}$ consisting of two different atomic orbitals $\ket{\varphi_0}$~and~$\ket{\varphi_1}$, i.e., 
\begin{equation}
\ket*{\phi_\mathrm{R}}=c_0\ket{\varphi_0}+c_1 \ket{\varphi_1}.
\end{equation}
The atomic orbitals $\ket{\varphi_0}$~and~$\ket{\varphi_1}$ are chosen to be real functions and the coefficients $c_\mathrm{0}$~and~$c_\mathrm{1}$ of the initially localized wave packet in this basis can also be chosen to be real numbers in order to avoid spurious probability currents between the two components.

We aim at maximizing the projection
\begin{equation}\label{eq:integration_range}
f(c_0,c_1)=-\frac{1}{\pi}\int_a^b \Im \ev*{G(E)}{\phi_\mathrm{R}}\dd{E}.
\end{equation}
Here, $G(E)$ is the Green's function averaged over the $\vb{k}$-points in the Brillouin zone and $a$~and~$b$ determine an energy range that encloses the resonance position.
We use the abbreviated notation
\begin{equation}
f(c_0,c_1)=\Im \left[\begin{pmatrix} c_0 \\ c_1 \end{pmatrix}^\dagger
\begin{pmatrix} M_{00} & M_{01} \\ M_{10} & M_{11} \end{pmatrix}
\begin{pmatrix} c_0 \\ c_1 \end{pmatrix} \right],
\end{equation}
where the matrix elements $M_{\mu\nu}$ are given by the complex values
\begin{equation}
M_{\mu\nu}=-\frac{1}{\pi}\int_a^b[SG(E)S]_{\mu\nu}\dd{E} \qquad \mathrm{for }\qquad \mu,\nu\in\{0,1\},
\end{equation}
and $S$ is the overlap matrix of the non-orthogonal basis set in \textsc{siesta}.

We optimize the above expression under the constraint $\braket*{\phi_\mathrm{R}}=c_0^2+c_1^2=1$ by substituting $c_1=\pm\sqrt{1-c_0^2}$ into $\tilde{f}(c_0)=f(c_0,\pm\sqrt{1-c_0^2})$ so that 
\begin{equation}\label{eq:f_tilde}
\tilde{f}(c_0)=\Im M_{11}+c_0^2 (\underbrace{\Im M_{00}-\Im M_{11}}_{=A})\pm c_0 \sqrt{1-c_0^2} (\underbrace{\Im M_{01} +\Im M_{10}}_{=B}).
\end{equation}

\begin{figure}[t]
\centering
\includegraphics[width=0.45\textwidth]{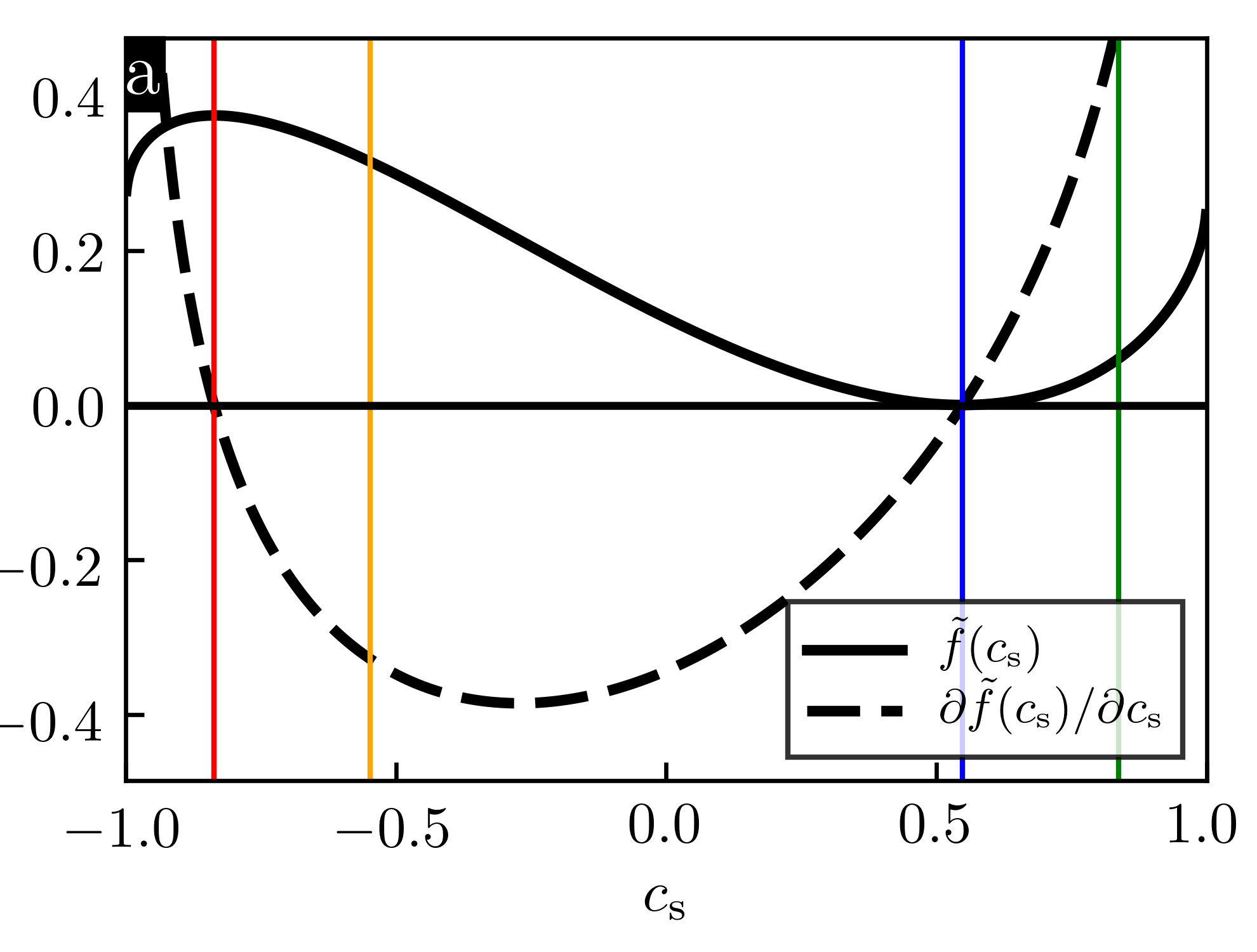}\includegraphics[width=0.45\textwidth]{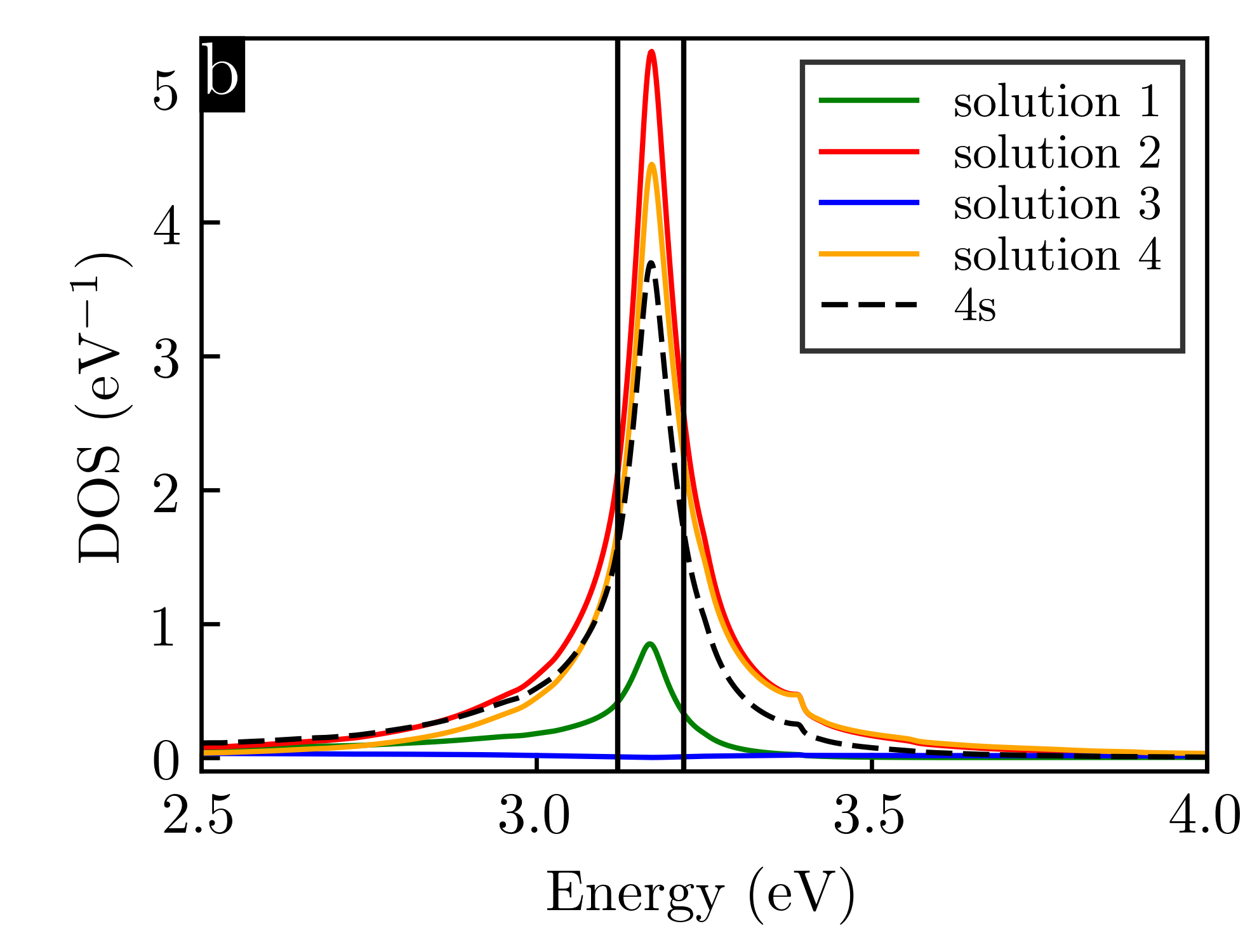}
 \caption{\label{fig:S1}Optimization of the resonance wave packet accounting for polarization in the vicinity of the surface. Example of the majority channel resonances of Ar$^*$ attached to Co(0001) at a distance of $3.6~${\AA}: (a) The four formal solutions of Eq.\;\ref{eq:biquadratic} (vertical colored lines) coincide in two cases with the roots of the derivative of $\tilde{f}(c_\mathrm{s})$ (Eq.\;\ref{eq:c_0_partial}, dashed line), where they maximize/minimize $\tilde{f}(c_\mathrm{s})$ (Eq.\;\ref{eq:f_tilde}). (b) Spectra corresponding to the four roots in subplot (a). Also shown in (b) is the spectrum of the pure Ar$^*$4s resonance (dashed black line). The vertical lines highlight the integration range from $a$~to~$b$ in Eq.\;\ref{eq:integration_range}.}
\end{figure}

Derivation with respect to the coefficient~$c_0$ yields
\begin{equation}\label{eq:c_0_partial}
\pdv{\tilde{f}}{c_0}=2c_0A\pm\frac{1-2c_0^2}{\sqrt{1-c_0^2}}B\stackrel{!}{=}0.
\end{equation}
Isolating the terms containing the coefficient~$c_0$ on one side of the equality and taking the square we receive
\begin{equation}
\frac{4c_0^2(1-c_0^2)}{(1-2c_0^2)^2}=\frac{B^2}{A^2}.
\end{equation}
This leads to the biquadratic equation
\begin{equation}\label{eq:biquadratic}
c_0^4-c_0^2+\frac{B^2}{4(A^2+B^2)}=0,
\end{equation}
which has four formal solutions
\begin{equation}\label{eq:4_solutions}
c_0=
\left\{ \begin{matrix}
\phantom{-}\sqrt{\frac{1}{2}+ \frac{1}{2} \sqrt{C}} \\
\phantom{-}\sqrt{\frac{1}{2}- \frac{1}{2} \sqrt{C}} \\
-\sqrt{\frac{1}{2}+ \frac{1}{2} \sqrt{C}} \\
-\sqrt{\frac{1}{2}- \frac{1}{2} \sqrt{C}}
\end{matrix} \right. \qquad \text{with} \qquad C=\frac{A^2}{A^2+B^2}.
\end{equation}
Of those solutions only two optimize the wave packet according to Eq.\;\ref{eq:c_0_partial}. 

\begin{table}[ht]
\caption{Extracted quantities corresponding to the majority and minority spin Ar$^*$-resonances on Co(0001). The tabulated values were obtained using either a pure Ar$^*$4s-orbital to represent the resonance wave packet $\ket{\phi_\mathrm{R}}$ or an optimized wave packet containing a 4p$_\mathrm{z}$-component. Shown are: the peak maximum $E_R$, the width obtained from a Lorentzian fitting $\Gamma$, the associated lifetime $\hbar/\Gamma$, and the lifetime $\tau_\mathrm{FT}$ derived using the alternative method described below. The percentages $|c_{\mathrm{p}}|^2$ of 4p$_\mathrm{z}$-contributions contained in the optimized wave packets are displayed in the last column.}
\centering
\begin{tabular}{@{}*{11}{P{0.06363\textwidth}@{}}}
\toprule
\multicolumn{2}{c}{Co(0001)}&\multicolumn{4}{c}{$\ket{\phi_{\mathrm{R}}}=\ket*{\varphi_{\mathrm{4s}}}$}&\multicolumn{5}{c}{$\ket{\phi_{\mathrm{R}}}=c_\mathrm{s}\ket*{\varphi_{\mathrm{4s}}} + c_\mathrm{p}\ket*{\varphi_{\mathrm{4p}_\mathrm{z}}}$}\\
\cmidrule(r){1-2}\cmidrule(lr){3-6}\cmidrule(l){7-11}
\multicolumn{2}{c}{Height}&$E_\mathrm{R}^{\mathrm{4s}}$&$\Gamma^{\mathrm{4s}}$&$\hbar/\Gamma^{\mathrm{4s}}$&$\tau_\mathrm{FT}^{\mathrm{4s}}$&$E_\mathrm{R}$&$\Gamma$&$\hbar/\Gamma$&$\tau_\mathrm{FT}$&$|c_\mathrm{p}|^2$\\
\multicolumn{2}{c}{(\r{A})}&(eV)&(meV)&(fs)&(fs)&(eV)&(meV)&(fs)&(fs)&(\%)\\
\midrule
\multirow{ 5}{*}{\rotatebox[origin=c]{90}{Majority}}
          &       2.4&      3.92&       141&      4.64&      5.38&      3.92&       128&      5.10&      5.67&        60\\
          &       2.7&      3.74&       139&      4.72&      4.74&      3.74&       137&      4.79&      4.85&        52\\
          &       3.0&      3.52&       116&      5.64&      5.71&      3.52&       118&      5.57&      5.74&        43\\
          &       3.3&      3.35&        97&      6.73&      6.60&      3.35&        99&      6.63&      6.63&        37\\
          &       3.6&      3.17&        98&      6.68&      6.27&      3.17&        97&      6.74&      6.28&        30\\
\cmidrule(r){1-2}\cmidrule(lr){3-6}\cmidrule(l){7-11}
\multirow{ 5}{*}{\rotatebox[origin=c]{90}{Minority}}
          &       2.4&      4.05&       253&      2.60&      2.81&      4.05&       219&      3.00&      3.33&        63\\
          &       2.7&      3.86&       200&      3.29&      3.48&      3.86&       179&      3.68&      4.06&        56\\
          &       3.0&      3.68&       182&      3.60&      4.18&      3.68&       155&      4.23&      5.08&        50\\
          &       3.3&      3.50&       226&      2.90&      4.43&      3.50&       181&      3.62&      4.86&        44\\
          &       3.6&      3.34&       308&      2.14&      3.92&      3.34&       251&      2.61&      4.63&        38\\

\bottomrule
\end{tabular}
\label{tab:co}
\caption{Extracted quantities corresponding to the majority and minority spin Ar$^*$-resonances on Fe(110). The values are defined as in Tab.\;\ref{tab:co}.}
\centering
\begin{tabular}{@{}*{11}{P{0.06363\textwidth}@{}}}
\toprule
\multicolumn{2}{c}{Fe(110)}&\multicolumn{4}{c}{$\ket{\phi_{\mathrm{R}}}=\ket*{\varphi_{\mathrm{4s}}}$}&\multicolumn{5}{c}{$\ket{\phi_{\mathrm{R}}}=c_\mathrm{s}\ket*{\varphi_{\mathrm{4s}}} + c_\mathrm{p}\ket*{\varphi_{\mathrm{4p}_\mathrm{z}}}$}\\
\cmidrule(r){1-2}\cmidrule(lr){3-6}\cmidrule(l){7-11}
\multicolumn{2}{c}{Height}&$E_\mathrm{R}^{\mathrm{4s}}$&$\Gamma^{\mathrm{4s}}$&$\hbar/\Gamma^{\mathrm{4s}}$&$\tau_\mathrm{FT}^{\mathrm{4s}}$&$E_\mathrm{R}$&$\Gamma$&$\hbar/\Gamma$&$\tau_\mathrm{FT}$&$|c_\mathrm{p}|^2$\\
\multicolumn{2}{c}{(\r{A})}&(eV)&(meV)&(fs)&(fs)&(eV)&(meV)&(fs)&(fs)&(\%)\\
\midrule
\multirow{ 5}{*}{\rotatebox[origin=c]{90}{Majority}}
          &       2.4&      3.67&       305&      2.16&      2.24&      3.67&       262&      2.51&      2.50&        63\\
          &       2.7&      3.53&       216&      3.05&      3.31&      3.54&       186&      3.53&      3.62&        56\\
          &       3.0&      3.37&       244&      2.70&      3.18&      3.37&       206&      3.18&      3.50&        50\\
          &       3.3&      3.21&       267&      2.46&      2.54&      3.21&       239&      2.74&      2.77&        43\\
          &       3.6&      2.98&       268&      2.45&      2.43&      2.99&       247&      2.66&      2.58&        33\\
\cmidrule(r){1-2}\cmidrule(lr){3-6}\cmidrule(l){7-11}
\multirow{ 5}{*}{\rotatebox[origin=c]{90}{Minority}}
          &       2.4&      3.74&      1209&      0.54&      0.78&      3.81&       740&      0.89&      1.00&        65\\
          &       2.7&      3.60&       736&      0.89&      1.01&      3.64&       590&      1.12&      1.19&        57\\
          &       3.0&      3.44&       571&      1.15&      1.21&      3.47&       495&      1.33&      1.34&        49\\
          &       3.3&      3.26&       509&      1.29&      1.33&      3.29&       456&      1.44&      1.43&        41\\
          &       3.6&      3.06&       472&      1.39&      1.44&      3.07&       444&      1.48&      1.49&        32\\
\bottomrule
\end{tabular}
\label{tab:fe}
\end{table}

To illustrate the construction of an optimized wave packet as described above we inspect the case of Ar$^*$ attached to Co(0001) at $3.6~${\AA} adsorption height. We consider a mixture of 4s (first $\zeta$ component of the \textsc{siesta} basis) and 4p$_\mathrm{z}$ components (as in the main text) and analyze the majority channel.

Fig.\;\ref{fig:S1}a shows the function $\tilde{f}(c_\mathrm{s})$ (cf. Eq.\;\ref{eq:f_tilde}) and its derivative $\partial \tilde{f}(c_\mathrm{s}) / \partial c_\mathrm{s}$ (cf. Eq.\;\ref{eq:c_0_partial}) in dependence of the weight $c_\mathrm{s}$ of the 4s-symmetry component. The colored vertical lines indicate the four formal solutions in Eq.\;\ref{eq:4_solutions}. Solution\;2 (red line) maximizes and solution\;3 (blue line) minimizes the function $\tilde{f}(c_\mathrm{s})$.

The Ar$^*$-resonances related to the four solutions in Fig.\;\ref{fig:S1}a are displayed in corresponding colors in Fig.\;\ref{fig:S1}b. Fig.\;\ref{fig:S1}b shows that solution\;2 (red line) visibly maximizes the area below the graph in the integration range from $a$~to~$b$ [cf. Eq.\;\ref{eq:integration_range}]. The integration range is marked by vertical black lines and is chosen to be $[E_\mathrm{R}-\Gamma/2;E_\mathrm{R}+\Gamma/2]$, where $E_\mathrm{R}$ and $\Gamma$ are the energy position $E_\mathrm{R}$ and the linewidth $\Gamma$ of the pure Ar$^*$4s-resonance. Clearly, the spectral feature of the optimized resonance state shows an increase in spectral density around the peak position in comparison with the pure Ar$^*$4s~resonance state (dashed black line).

\begin{figure}[t]
\includegraphics[width=0.45\linewidth]{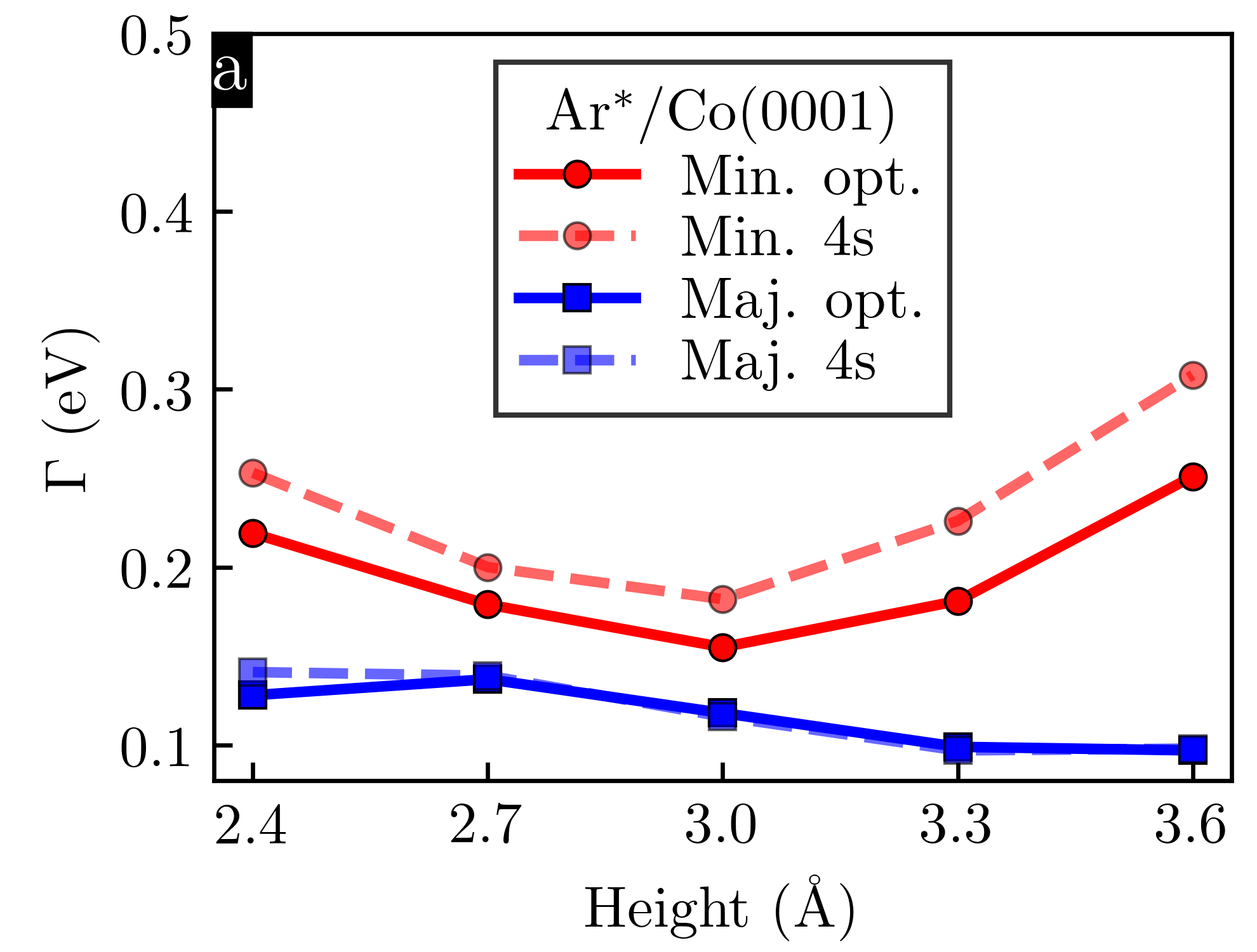}\includegraphics[width=0.45\linewidth]{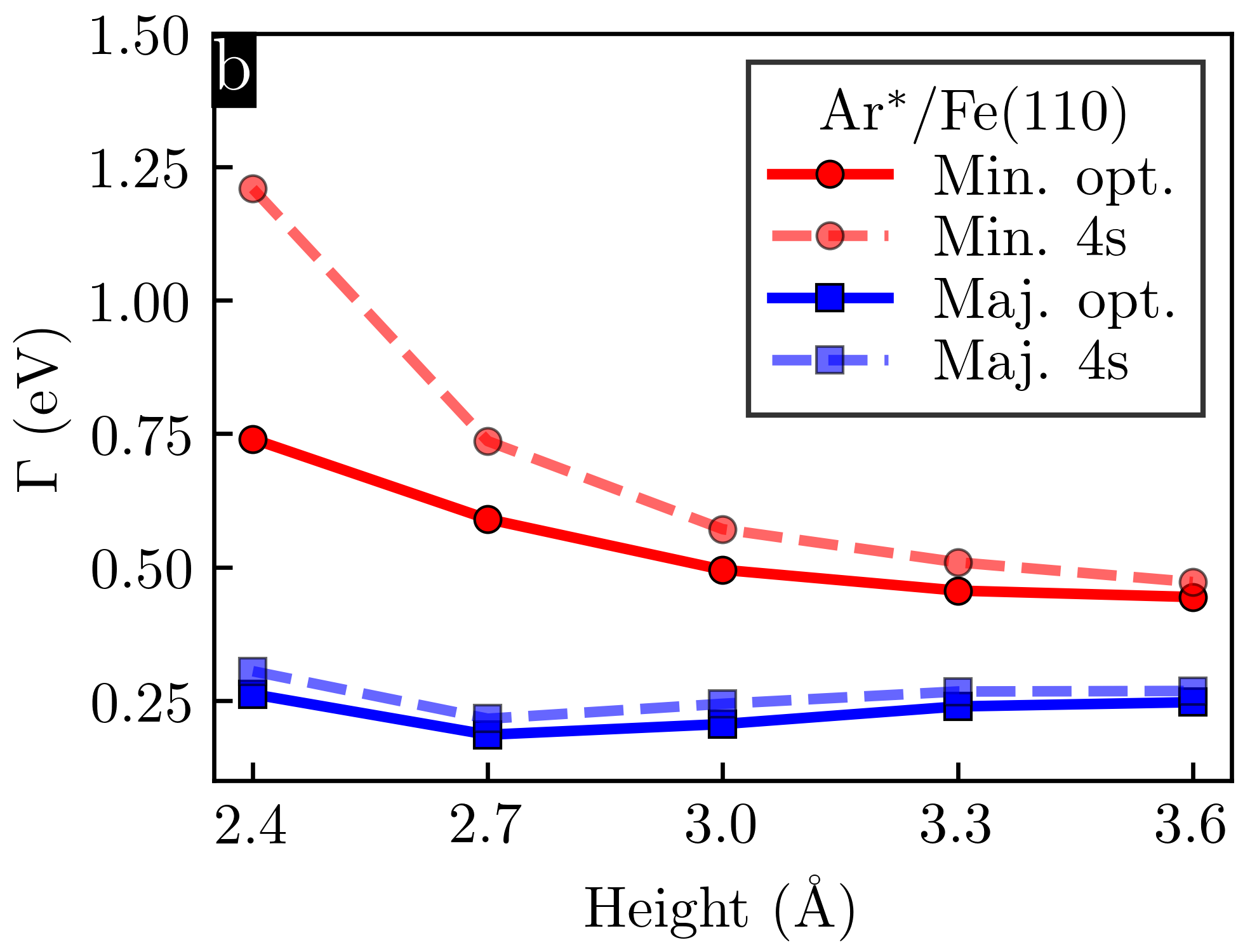}
\caption{Extracted linewidths of Ar$^*$ resonances with respect to different adsorption heights after Lorentzian fitting.
Minority (red lines) and majority channels (blue lines) are displayed for (a) Ar$^*$/Co(0001) and (b) Ar$^*$/Fe(110). Continuous lines refer to values obtained from optimized wave packets (opt.) accounting for polarized Ar$^*$-resonances, while dashed lines refer to values related to pure Ar$^*$4s-resonances.}
\label{fig:S2}
\end{figure}

Extracting the widths of the optimized spectra containing a polarizing component we obtain consistently larger lifetimes than for pure Ar$^*$4s-resonances. All extracted numerical values of the optimized spectra are summarized in Tab.\;\ref{tab:co} and Tab.\;\ref{tab:fe}. Listed are the positions of the maxima of the resonances $E_\mathrm{R}$, the linewidths $\Gamma$ obtained from a Lorentzian fitting, and the corresponding lifetimes $\hbar/\Gamma$. These values can be compared with the values related to pure Ar$^*$4s-resonances. For a visual comparison, Fig.\;\ref{fig:S2} additionally shows the extracted linewidths of optimized (opt.) and pure 4s-resonances for Ar/Co(0001) (Fig.\;\ref{fig:S2}a) and for Ar/Fe(110) (Fig.\;\ref{fig:S2}b). The optimized resonances exhibit a reduced linewidth, while the overall trends remain unchanged upon inclusion of the polarizing p$_\mathrm{z}$-component.

\begin{figure}[ht]
\includegraphics[width=0.9\linewidth]{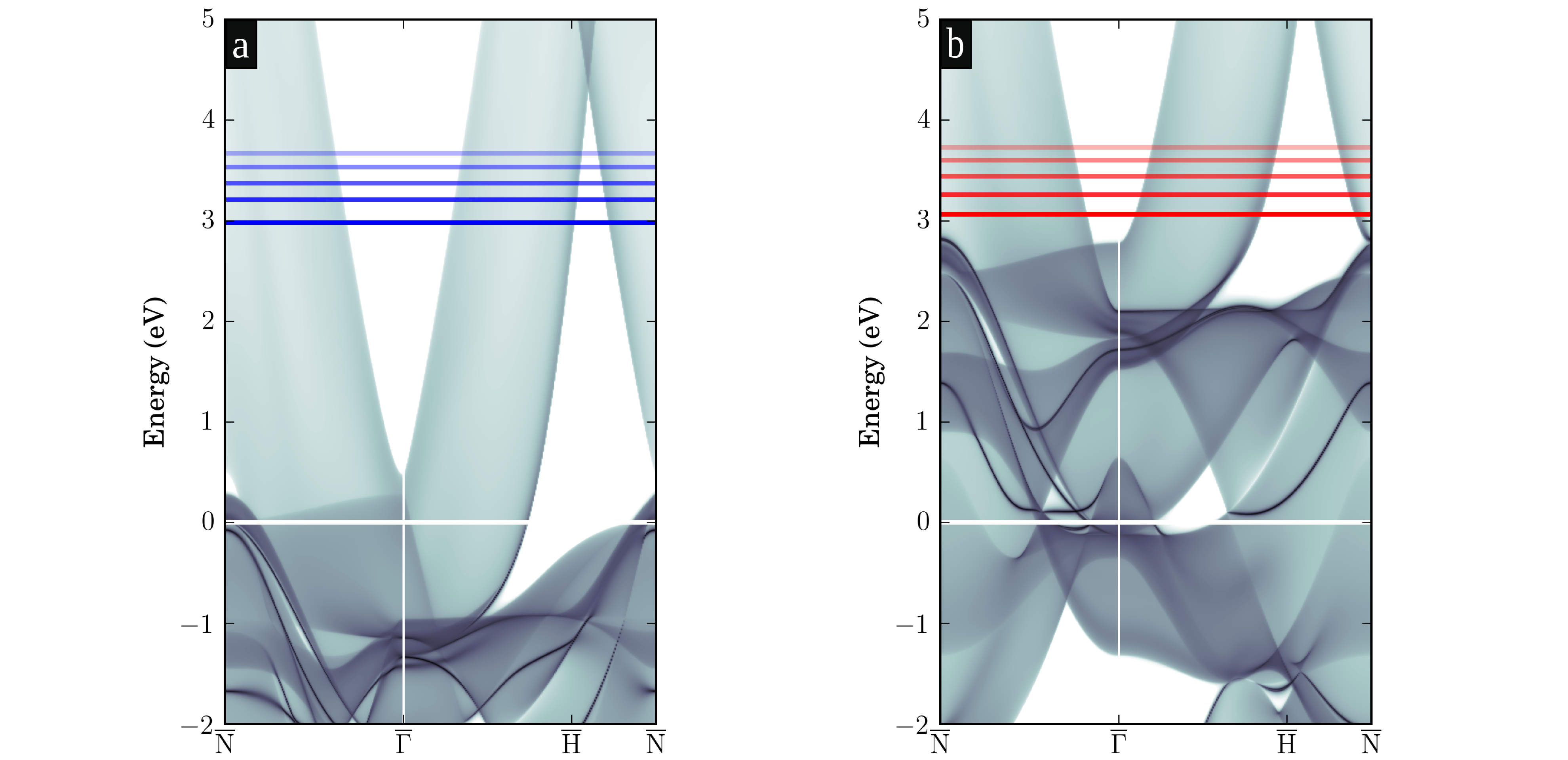}
\caption{Energy positions of Ar$^*$ resonances with respect to the band structure of clean Fe(110) surfaces. Majority (blue lines) and minority channels (red lines) are displayed in (a) and (b), respectively.}
\label{fig:S3}
\end{figure}

As in the case of Ar/Co(0001) the observed spin dependence in the lifetimes for Ar/Fe(110) is rooted in the increased sizes (in reciprocal space) of the electronic gaps around the $\overline{\Gamma}$-point in the majority channel. The alignment of the resonance positions with respect to the surface band structure of Fe(110) is shown in Fig.\;\ref{fig:S3}.

The last columns in Tab.\;\ref{tab:co} and Tab.\;\ref{tab:fe} show the percentages\;$|c_\mathrm{p}|^2$ of the polarizing component contained in the optimized resonance wave packets. As one expects, the polarization of the wave packets increases when lowering the distance between the Ar$^*$ atoms and the surface. 

We additionally employed an alternative scheme to extract the lifetime $\tau_\mathrm{FT}$ after Fourier transformation (FT) to the time-domain (cf. $\tau_\mathrm{FT}$ in Tab.\;\ref{tab:co} and Tab.\;\ref{tab:fe}). The scheme is described in the section below. Looking at the values $\tau_\mathrm{FT}$ in the tables and comparing them to the values from Lorentzian fittings $\hbar/\Gamma$ one can see that overall the trends remain unchanged. 
However, in the case of Ar$^*$/Co(0001) at a distance of $3.3$\;\r{A} and $3.6$\;\r{A} growing asymmetries in the line shapes of the resonances bring about deviations from a symmetric Lorentzian fitting in the minority channel.  

\section{Extraction of Lifetimes in the time-domain}
An alternative to the extraction of lifetimes by Lorentzian fittings of resonance features, is to compute the survival amplitude $A(t)$ associated with a given resonance. The survival amplitude $A(t)$ of a state $\phi(t)$ at a time $t>t_0$ is given by 
\begin{equation}\label{eq:amplitude}
A(t)=\braket{\phi(t_0)}{\phi(t)}=\ev{\mathrm{i}G(t-t_0)}{\phi(t_0)},
\end{equation}
where the Green's function $G(t-t_0)$ acts as propagator in the time-domain. 
Once the survival amplitude is known, the survival probability $S(t)=|A(t)|^2$ can be obtained and the mean lifetime $\tau_\mathrm{FT}$ can be computed by
\begin{equation}\label{eq:tau}
\tau_\mathrm{FT}=\frac{\int_0^\infty t\, S(t) \dd{t}}{\int_0^\infty S(t) \dd{t}}\qquad\mathrm{with}\qquad S(t)=|A(t)|^2.
\end{equation}

The Green's function in the time-domain $G(t-t_0)$ is given by the Fourier transform of the Green's function $\tilde{G}(E)$ associated with the resonance peak, 
\begin{equation}\label{eq:fourier}
G(t-t_0)=\frac{1}{2\pi}\int_{-\infty}^\infty \tilde{G}(E) \mathrm{e}^{-\mathrm{i}E(t-t_0)}\dd{E}.
\end{equation}

In order to define the Green's function $\tilde{G}(E)$ associated with the resonance we crop the spectrum above the Fermi level using a Fermi distribution $f(E)$ (only unoccupied states are available for the propagation of the adsorbate's excited electron).  
\begin{equation}
\Im[\tilde{G}(E)]=[1-f(E)]\Im[G(E)],
\end{equation}
where the imaginary part of the Green's function\;$G(E)$ is related to the spectral density via $\varrho(E)_\mathrm{R}=-1/\pi\Im[G(E)]$.
Finally, the real part of the Green's function related to the cropped spectrum is obtained by using the Kramers-Kronig relation
\begin{equation}
\Re[\tilde{G}(E)]=\frac{1}{\pi}\operatorname{P}\int_{-\infty}^\infty\frac{\Im[\tilde{G}(E')]}{E'-E}\dd{E'}.
\end{equation}
Knowing the real and imaginary part of the Green's function associated with the resonance, a Fourier transformation to the time-domain (Eq.\;\ref{eq:fourier}) yields the survival amplitude (Eq.\;\ref{eq:amplitude}) and thus the mean lifetime $\tau_\mathrm{FT}$ (Eq.\;\ref{eq:tau}). In practice, we evaluated Eq.\;\ref{eq:tau} by numerical integration up to $500$\;fs. 

\section{Site Dependence of the Linewidths of the Resonances}
To confirm that the observed trends in this work are independent of the placement of the Ar$^*$ atoms on the Co(0001) and Fe(110) surfaces, we investigated the variation of the resonant linewidths across different adsorption sites. The extracted linewidths of pure 4s\;resonances are listed in Tab.\;\ref{tab:sites}. Inspecting the values no site dependence of the linewidths~$\Gamma$ is found apart from weak fluctuations (less than $14$~meV).

\begin{table}[h]
\caption{Extracted linewidths~$\Gamma$ in meV for Ar$^*$4s resonances with respect to different adsorption sites of Argon on Fe(110) and Co(0001):  top~position (tp), short~bridge (sb), long~bridge (lb), bridge~position (br), fcc~hollow (fcc), and hcp~hollow (hcp). Shown are values for different adsorption heights~$h$ in~{\AA}. }
\centering
\begin{tabular}{@{}*{11}{P{0.05555\textwidth}@{}}}
\toprule
&&\multicolumn{3}{c}{Fe(110)}&\multicolumn{4}{c}{Co(0001)}\\
\cmidrule(r){3-5}\cmidrule(l){6-9}
&$h$&tp&sb&lb&tp&br&fcc&hcp\\
\midrule
\multirow{ 3}{*}{\rotatebox[origin=c]{90}{Maj.}}
          &       2.7&      216&       215&      212&      139&      143&     143&       144\\
          &       3.0&      244&       247&      248&      120&      128&     128&       128\\
          &       3.0&      267&       273&      275&      102&      107&     108&       108\\
\midrule
\multirow{ 3}{*}{\rotatebox[origin=c]{90}{Min.}}
          &       2.7&      736&       734&      727&      189&      194&       195&     195\\
          &       3.0&      571&       584&      585&      178&      191&       190&     191\\
          &       3.3&      509&       518&      520&      234&      247&       247&     248\\
\bottomrule
\end{tabular}
\label{tab:sites}
\end{table}

\newpage
\section{Extracted Lifetimes for Argon on Nickel}
\begin{figure}[h]
\includegraphics[width=0.45\linewidth]{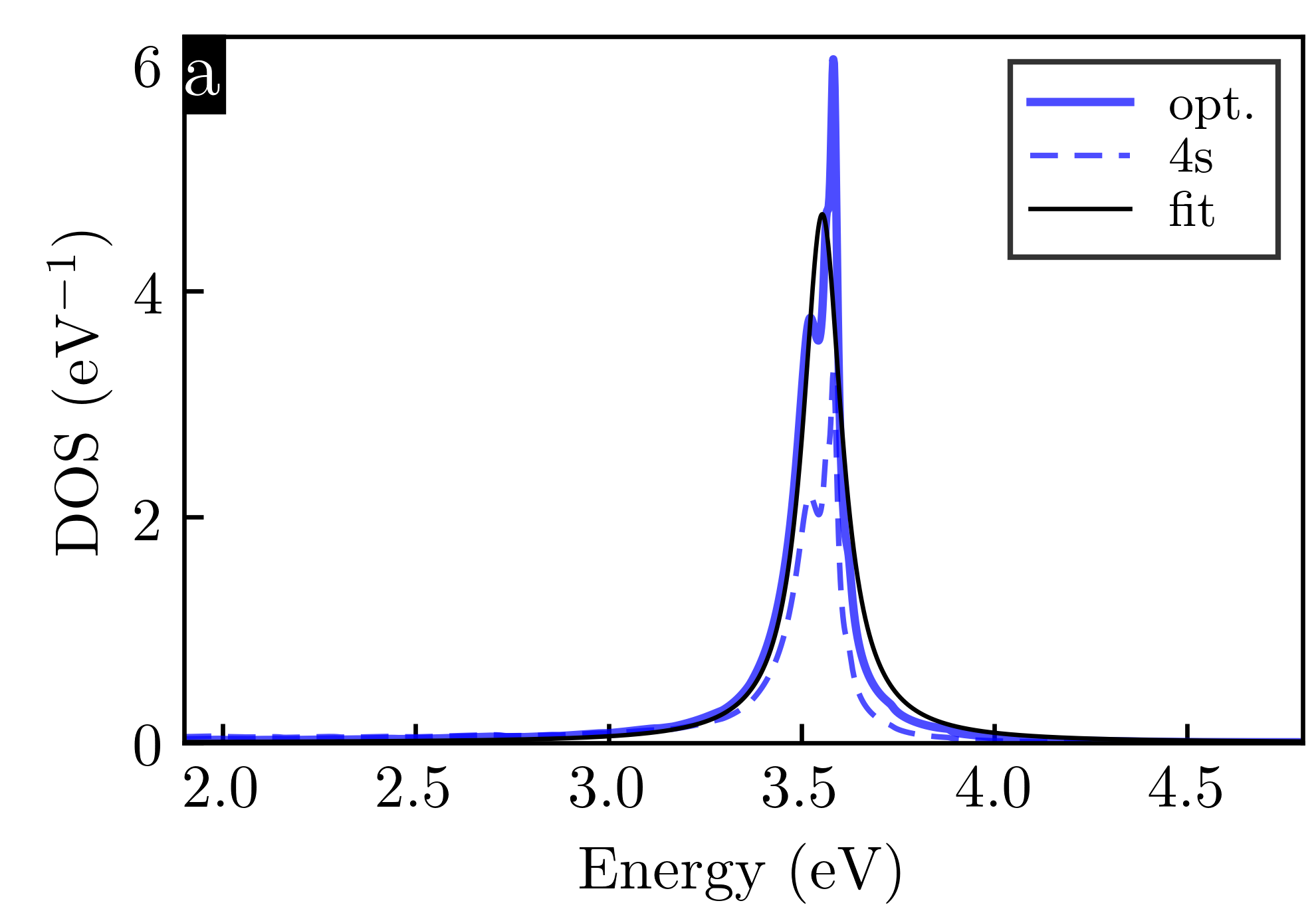}\includegraphics[width=0.45\linewidth]{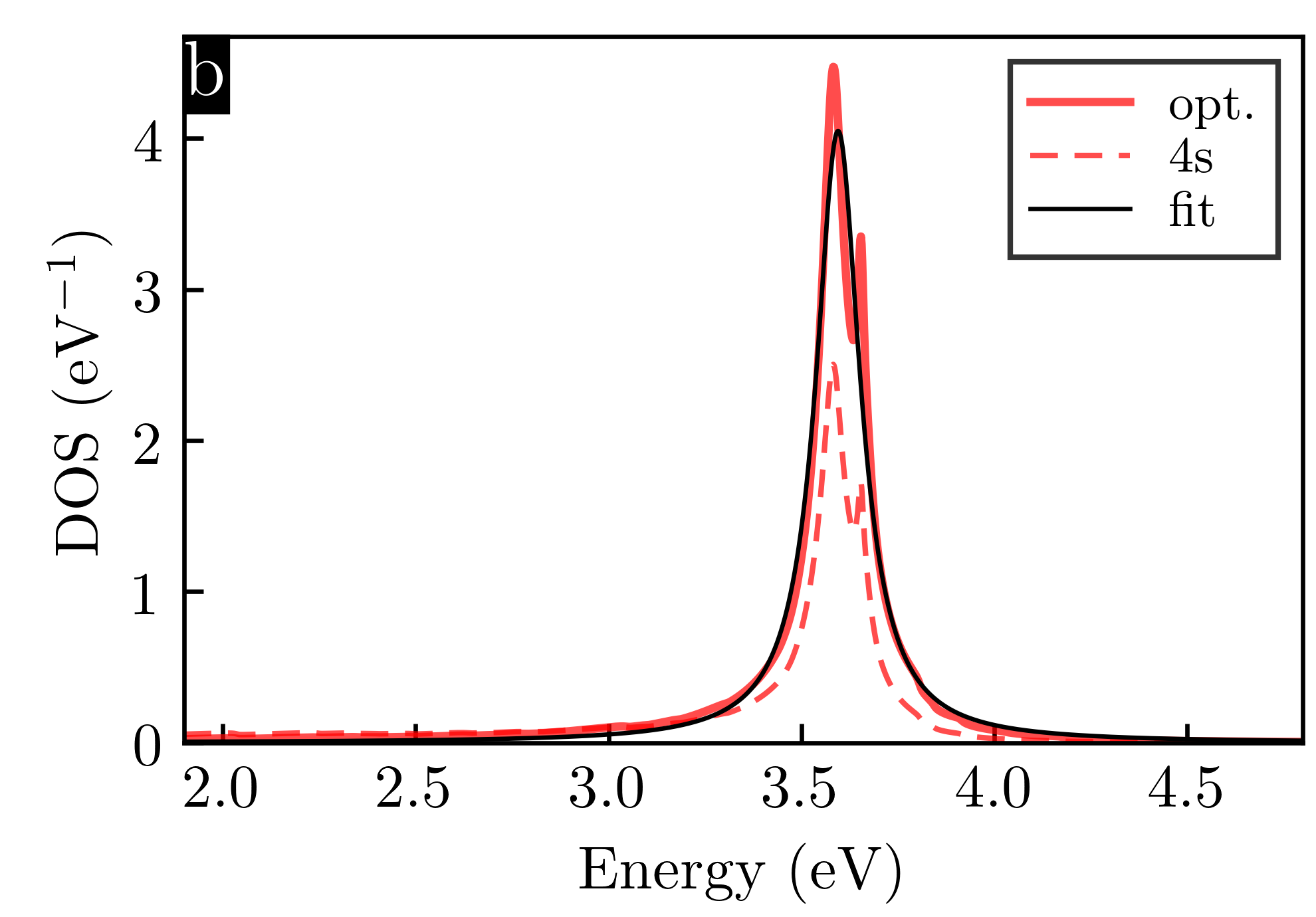}
\caption{Extracted line shapes of Ar$^*$ resonances at Ni(111) surfaces. Majority (a) and minority (b) spin channel. Dashed lines refer to pure Ar$^*$4s resonance wave packets (4s) and continuous lines to optimized resonance wave packets (opt.) accounting for polarization. The continuous black lines represent Lorentzian fits of the optimized resonance shapes.}
\label{fig:S4}
\end{figure}

In analogy to the experiments by Blobner et al.~\cite{Blobner2014}, we also investigated the lifetimes of Ar$^*$ resonances on Ni(111) at an adsorption height of $3.0$\;\r{A}. We initially modeled the system using a symmetric slab consisting of 13 layers. In a second step we computed the Green's function of the first 9 layers of which the last 6 were connected via self-energy terms (Eq.\;2 of the main text) to an infinite bulk substrate. We used a 6$\times$6$\times$1 $\vb{k}$-point sampling. The remaining computational settings were equal to the calculations for Ar$^*$ on Fe(110) or Ar$^*$ on Co(0001) and are reported in the main text. 
The resulting resonance lineshapes for pure 4s and polarized (opt.) wave packets are depicted in Fig.\;\ref{fig:S4} from which we extracted the values of the energy positions\;$E_\mathrm{R}$ ($3.58$\;eV for both spins), linewidths\;$\Gamma$, and lifetimes $\hbar/\Gamma$ associated with the Ar$^*$ resonances.

For Ar/Ni(111) the charge transfer times $\hbar/\Gamma$ differ by only 14\% in both spin channels using polarized resonance wave packets (opt.), the majority channel showing a charge transfer time of $\hbar/\Gamma=5.07$\;fs ($\Gamma=129$\;meV) versus $\hbar/\Gamma=5.79$\;fs ($\Gamma=113$\;meV) in the minority channel. Pure Ar$^*$4s-resonances lead to similar results with $\hbar/\Gamma^{\mathrm{4s}}=5.45$ ($\Gamma^{\mathrm{4s}}=120$\;meV) in the majority and $\hbar/\Gamma^{\mathrm{4s}}=5.13$ ($\Gamma^{\mathrm{4s}}=128$\;meV) in the minority channel.  
Therefore, our calculations result in very faint differences for charge transfer in both spin channels reflecting a reduced spin splitting of the bands of the magnetic substrate and approaching a situation of equally fast charge transfer in both spin channels as reported experimentally for Ni(111) surfaces~\cite{Blobner2014}.

Moreover, these results contrast our calculations of $\hbar/\Gamma$ on Ar/Co(0001) and Ar/Fe(110), where we found pronounced differences with respect to charge transfer in both spin channels, see Tab.\;\ref{tab:co}. Looking at Tab.\;\ref{tab:co} we find that the calculated charge transfer times on Ni(111) also come close to the theoretically determined time-scale for majority channel charge transfer of $5.57$\;fs on Co(0001) at an adsorption distance of $3.0$\;\r{A}. This observation resembles the experimentally reported overall behavior across different substrates~\cite{Blobner2014}. 

\bibliography{supplemental_material}